\documentclass[prd,tightenlines,nofootinbib,twocolumn,superscriptaddress]{revtex4}

\usepackage{amsfonts,amssymb,amsthm,bbm}

\usepackage{amsmath}

\usepackage{hyperref}

\usepackage{color,psfrag}
\usepackage[dvips]{graphicx}

\usepackage{dsfont}

\def\bef{\begin{figure}}
\def\eef{\end{figure}}

\newcommand{\eeqn}{\end{eqnarray}}

\newcommand{\eps}{\varepsilon}

\def\la{\lambda}

\def\be{\begin{equation}}
\def\ee{\end{equation}}
\def\bea{\begin{eqnarray}}
\def\eea{\end{eqnarray}}
\def\beq{\begin{eqnarray}}
\def\eeq{\end{eqnarray}}
\def\bas{\begin{subequations}\begin{eqnarray}}
\def\eas{\end{eqnarray}\end{subequations}}

\def\nn{\nonumber}

\def\la{\langle}
\def\ra{\rangle}
\def\f{\frac}

\def\SL{\text{SL}}

\newcommand{\R}{{\mathbb R}}
\newcommand{\Z}{{\mathbb Z}}

\newcommand{\cH}{{\mathcal H}}

\newcommand{\cV}{{\mathcal V}}

\newcommand{\cC}{{\mathcal C}}
\newcommand{\cS}{{\mathcal S}}

\def\pp{\partial}
\def\rd{\textrm{d}}

\def\ka{\kappa}

\def\eps{\epsilon}

\def\la{\langle}
\def\ra{\rangle}

\newcommand{\bes}{\begin{eqnarray}}
\newcommand{\ees}{\end{eqnarray}}

\renewcommand{\sl}{{\mathfrak{sl}}}

\def\pp{\partial}

\def\ka{\kappa}

\def\eps{\epsilon}

\begin{document}

\title{Polymer  Quantum Cosmology: \vspace*{1mm}\\ Lifting quantization ambiguities using a $\SL(2,\mathbb{R})$ conformal symmetry}

\author{{ J. Ben Achour}}\email{jibrilbenachour@gmail.com}
\affiliation{Center for Relativity and Gravitation, Beijing Normal University, Beijing 100875, China}

\author{Etera R. Livine}
\email{etera.livine@ens-lyon.fr}
\affiliation{Perimeter Institute for Theoretical Physics, 31 Caroline Street North, Waterloo, Ontario, Canada N2L 2Y5}
\affiliation{Universit\'e de Lyon, ENS de Lyon,  Laboratoire de Physique, CNRS UMR 5672, F-69342 Lyon, France}

\begin{abstract}

In this letter, we stress that the simplest cosmological model consisting in a massless scalar field minimally coupled to homogeneous and isotropic gravity has an in-built $\SL(2,\mathbb{R})$ symmetry. Protecting this symmetry naturally provides an efficient way to constrain the quantization of this cosmological system whatever the quantization scheme and allows in particular to fix the quantization ambiguities arising in the canonical quantization program.
Applying this method to the loop quantization of the FLRW cosmology leads to a new loop quantum cosmology  model which preserves the  $\SL(2,\mathbb{R})$ symmetry of the classical system. This new polymer regularization consistent with the conformal symmetry can be derived as a non-linear canonical transformation of the classical FLRW phase space, which maps the classical singular dynamics into a regular effective bouncing dynamics. This improved regularization preserves the scaling properties of the volume and Hamiltonian constraint. 3d scale transformations, generated by the dilatation operator, are realized as unitary transformations despite the minimal length scale hardcoded in the theory. Finally, we point out that the resulting cosmological dynamics  exhibits an interesting duality between short and long distances, reminiscent of the T-duality in string theory, with the near-singularity regime dual to the  semi-classical regime at large volume.
The technical details of the construction of this model are presented in a longer companion paper \cite{BenAchour:2019ywl}.

\end{abstract}


\maketitle

Finding a consistent and ambiguity-free realization of canonical quantum cosmology remains a challenging task. Since the first approach by Wheeler and De Witt, numerous efforts have been devoted to refining the canonical construction in order to build a quantum theory of the universe as a single homogeneous object. However, it remains hard to extract unambiguous predictions for early cosmology from the different quantizations schemes. See \cite{Wiltshire:1995vk, Bojowald:2010cj, Bojowald:2015iga} for reviews. In the canonical framework, the resulting quantum dynamics usually suffers from a large degree of arbitrariness since the construction of the quantum theory relies on several choices, such as the inner product and the operator ordering  in  the Hamiltonian constraint \cite{Steigl:2005fk, Bojowald:2014ija, Livine:2012mh}, but also  appropriate boundary conditions to select particular physical solutions for the wave function. Moreover, in the canonical program,
the definition of the quantum states and their dynamics and the resulting phenomenology crucially depend on the choice of slicing of the 4d geometry. This is true for the standard Wheeler-De Witt quantization, based on the Schr\"odinger representation, but also for the polymer (or loop) quantization scheme (see \cite{ Corichi:2006qf, Fredenhagen:2006wp, Corichi:2007tf} for self-contained presentations of the polymer quantization). 

The polymer quantization framework, which captures the lattice-like structure of the geometry inherited from loop quantum gravity, is based on a different set of basics operators which are non-linear in the original canonical variables and requires a regularization of the Hamiltonian scalar constraint. This  introduces an extra layer of ambiguities, especially in the choice of regularization scheme (usually referred to as the choice of connection versus curvature schemes) and the choice of the spin used when representing the curvature entering the Hamiltonian constraint in terms of holonomy operators. While the standard choices employed in the literature can be justified by physical motivations, these quantization ambiguities are nonetheless inherent to the polymer quantization program and  drastically affect the resulting quantum dynamics of the universe. See \cite{BenAchour:2016ajk} for a systematic investigation of these ambiguities in flat and isotropic loop quantum cosmology, and \cite{Singh:2013ava} for anisotropic models as well as \cite{Corichi:2011pg, Dupuy:2016upu} for the closed universe case. At the end of the day, it is crucial to distinguish the predictions which do not depend on such ambiguities from those which remain sensitive to such choices.

Finally, another issue appears due to intertwined role of the generator of 3d scale transformations and the generator of the deparametrized dynamics. For instance, in the context of a massless scalar field minimally coupled to flat FLRW universe, these two observables coincide and the cosmological dynamics in term of the scalar field time is nothing else than a field-dependent rescaling of the induced metric and the extrinsic curvature of the spatial 3d slice. This means that having a well-defined unitary cosmological evolution is equivalent to having a well-defined unitary realization of scale transformations at the quantum level.
Now the polymer quantization's key ingredient is  a minimal length scale which signals the quantum gravity regime. This new length scale plays the role of a universal cut-off of the quantum theory and should therefore be unaffected under global rescaling of the geometry. In standard LQC, this question has not been addressed yet and the area gap, which encodes the minimal length scale, fails to be invariant under global rescaling. This is usually referred to as the ``Immirzi ambiguity''. It implies in particular that  the standard LQC construction does not propose a consistent implementation of  scale transformations at the quantum level and that their relation to the cosmological evolution is broken.

\medskip

In this letter, we present a new path to address some of the ambiguities discussed above. More precisely, we will show how to construct a new consistent LQC model with unitary scale transformations. We focus on the inherent $\SL(2,\R)$ symmetry of classical cosmology and introduce the new, though natural, criteria of preserving this symmetry at the quantum level. This fixes the factor ordering ambiguities at the quantum level and allows for a well-defined Hermitian dilatation generator and thus for unitary scale transformations. While the present work focuses on applying this method of protecting the  conformal symmetry in the homogeneous and isotropic FLRW cosmology, we stress that the $\SL(2,\R)$ structure also holds for more general models, such as the Bianchi I cosmology \cite{BenAchour:2019ywl}.

This $\SL(2,\mathbb{R})$ symmetry of the cosmological phase space  describes the scaling properties of the Hamiltonian constraint and the physical volume of the universe, and thus encodes the full cosmological dynamics. By preserving this conformal structure up in the quantum theory, it allows to recast LQC as a $\SL(2,\mathbb{R})$-invariant quantum cosmology with a universal minimal length scale.
Besides  allowing to entirely  solve the theory solely based on symmetry arguments, it opens new directions of research, such as the possibility of bootstrapping quantum cosmology or a mapping to the conformal quantum mechanics of de Alfaro, Fubini and Furland (dAFF) \cite{deAlfaro:1976vlx}.

This letter is intended to provide a  summary of the construction of the new  LQC theory at both classical and quantum levels and present its key ingredients and features, while technical details  appear in a longer companion paper \cite{BenAchour:2019ywl}.

\section{Classical $\sl(2,\mathbb{R})$ structure}

Consider the FLRW line element given by
\be
\rd s^2 = - N^2(\tau) \rd\tau^2 + a^2(\tau)\delta_{ab} \rd x^a \rd x^b
\ee
where $N(\tau)$ the homogeneous lapse function and $a(\tau)$ is the scale factor.
We consider the simplest cosmological model consisting in a massless homogeneous scalar field $\phi(\tau)$ minimally coupled to this metric. The symmetry reduced action of this system is given by
\begin{align}
\cS[a, N, \phi] & = V_{\circ} \int dt \left[ - \frac{3}{8\pi G} \frac{a \dot{a}^2}{N} + \frac{a^3}{2N} \dot{\phi}^2 \right]
\,,
\end{align}
where we have integrated over a fiducial 3d cell of co-moving volume $V_\circ$ .
A useful set of canonical variables for the gravitational sector is given by the 3d volume $v$ of the fiducial cell and  the extrinsic curvature (or Hubble rate) $b$, related to the scale factor $a$ and its first derivative $\dot{a}$ by:
\be
v = a^3 V_{\circ},
 \qquad
 b 
  = \frac1{4 \pi G} \frac{\dot{a}}{Na},
\ee 
As for the matter sector, we further introduce the conjugated momentum $\pi_{\phi}$ to the scalar field $\phi$,
\be
\pi_{\phi} = V_{\circ} \frac{a^3 \dot{\phi}}{N}
\,.
\ee 
They form pairs of canonically conjugate variables, $\{ b,v\} = 1$ and $\{ \phi, \pi_{\phi}\} = 1$.

We work in units where we have  set $\hbar = c = 1$ for the sake of simplicity. There are two natural length scales associated to this system: an infrared scale  $\ell_{\circ}={}^3\sqrt{V_{\circ}}$ set by the size of the fiducial cell, and a UV scale given by the Planck length $\ell_{\text{Planck}} \propto \sqrt{G}$.
Then the canonical variables have dimensions $ [v]= [b]^{-1} = L^3$ and $[\phi]=[\pi_{\phi}]^{-1}=L^{-1}$.

The action reads $\cS=\int \rd t\,\big{[} \dot{b}v+\dot{\phi}\pi_{\phi}-H[N]\big{]}$ with the Hamiltonian $H[N]$: 
\be
\label{Hclass}
H [N]
= NH= \f N2 \left(\frac{\pi^2_{\phi}}{ v} -\kappa^2 v b^2\right)
\,,
\ee
where  we have introduced the Planckian size length scale $\ka=\sqrt{12 \pi G}$. The lapse $N$ plays the role of a Lagrange multiplier enforcing the Hamiltonian scalar constraint $H=0$, which generates the invariance under time-reparametrization.
Since the  constraint does not depend on the scalar field $\phi$, its conjugate momentum is a constant of motion, i.e. $\dot{\pi}_{\phi} = 0$.

Beside the standard invariance under time-reparametrization, there is residual time-dependent conformal symmetry which can be seen as follow. %
Let us consider the integrated trace of the extrinsic curvature,
\be
\int d^3x \; \sqrt{|q|} q^{ab} K_{ab} = \ka^2\, C
\quad\textrm{with}\,\,
C=vb\,.
\ee
This observable generates  3d scale transformations of the space-like hypersurface in our slicing, 
\be
e^{\eta\{ C, \cdot\}} \triangleright v = e^{\eta} v \qquad e^{\eta\{ C, \cdot\}} \triangleright b = e^{-\eta} b
\ee
This flow rescales the 3d volume $v$ and the extrinsic curvature  $b$ with inverse factors, so that  $\cC$ generates a canonical transformation\footnotemark{}.
\footnotetext{
In the Ashtekar-Barbero variables, the extrinsic curvature $b$ is proportional to the Immirzi parameter, so the dilatation generator $\cC$ can be interpreted as rescaling of the Immirzi parameter. It is actually the complexifier in the usual vocabulary of loop quantum gravity \cite{BenAchour:2017qpb}. This underlines that one can rescale freely the Immirzi parameter at the classical level by a canonical transformation, thus without affecting the classical physics \cite{Rovelli:1997na}. This might nevertheless change if we also couple fermions to gravity, since the Immirzi parameter can be understood as controlling torsion degrees of freedom \cite{Perez:2005pm}.
}
Now, the crucial point is that this dilatation generator forms a closed Poisson algebra with the Hamiltonian of the system and the 3d volume, with the following brackets:
\begin{align}
 \{v, H\} &= \ka^2C\,, 
 \nn\\
 \{ C, H \} & = - H\,, 
 \label{23}\\
 \{ C, v \} &= v\,.\nn
\end{align}
This algebra, which we call the CVH algebra following \cite{BenAchour:2017qpb}, encodes the whole dynamics of the cosmological model. Indeed, it implies that $C$ is the speed of evolution of the volume $v$,
\be
\f{\rd v}{\rd \tau}=\{v,H[N]\}=N\ka^2C
\,,
\ee
while $C$ itself is a constant of motion,
\be
\f{\rd C}{\rd \tau}=\{C,H[N]\}=-H[N] \sim 0\,,
\ee
since the Hamiltonian vanishes on-shell.

This CVH algebra is actually isomorphic to the $\sl(2,\mathbb{R})$ Lie algebra. The identification to the  $\sl(2,\mathbb{R})$ algebra generators is done up to an arbitrary dimensionless constant\footnotemark{} $\sigma\in\R$:
\begin{align}
k_y & = C\,,\nn
\\
k_x & = \frac{1}{2\sigma \kappa^3} v +{\sigma \kappa} H \,,
\label{sl2CVH}\\
j_z & =  \frac{1}{2\sigma \kappa^3} v -{\sigma \kappa} H
\,,\nn
\end{align}
\footnotetext{
This is the same parameter appearing in the definition of the $\sl(2,\R)$ generators in conformal quantum mechanics \cite{deAlfaro:1976vlx}.
}
where the Planck length $\ka = 12\pi G$ is inserted to ensure that the objects all have the same physical dimension. These generators satisfy the standard $\sl(2,\mathbb{R})$ commutations relations 
\be
\label{sl2alg}
\{ k_x, k_y\} = - j_z 
\,,\,\,
\{ j_z, k_x\} = k_y
\,, \,\,
\{ j_z, k_y\} = - k_x
\,.
\ee
So the dilatation generator $C$ is a boost generator while the volume $v$ and the Hamiltonian constraint $H$ are null generators.
Finally, the Casimir of this conformal algebra coincides with the squared momentum of the scalar field, which is a constant of motion:
\be
\label{sl2Casimir}
\mathfrak{C_{\sl_{2}}} = j^2_z - k^2_x - k^2_y =\f2{\ka^2}vH - C^2
=
- \frac{\pi^2_{\phi}}{\ka^2}
\,.
\ee
This provides this simple cosmological model with an elegant group structure.  This hidden conformal invariance, also noticed in earlier work \cite{Pioline:2002qz}, can be used as a powerful new criteria to restrict the quantization ambiguities when performing a canonical quantization of the system, whatever particular  regularization and quantization scheme is used. Below, we propose to apply this criteria to the polymer quantization program at the root of loop quantum cosmology.

\section{New regularization preserving the $\sl(2,\mathbb{R})$ structure}

The polymer quantization does not quantize directly the extrinsic curvature $b$ as a fundamental operator of the quantum theory, but trades it for its exponentiated version $e^{i \lambda b}$, where $\lambda$ is a fundamental regularization scale, $[\lambda] = L^3$.
This length scale (or more precisely volume scale) is inherited from loop quantum gravity and is of Planck size:
\be
\lambda = \Delta \ell_{\text{Planck}}^3
\ee
where the dimensionless proportionality constant $\Delta$ is typically chosen from the area or volume gap of loop quantum gravity\footnotemark.  We can nevertheless define loop quantum cosmology without any reference to loop quantum gravity and its area and volume spectra. Then $\Delta$, and thus $\lambda$, remains a free unspecified constant, which should be determined a posteriori from comparing the theory predictions with experimental data.
\footnotetext{
In standard LQC, $\Delta$ is fixed by demanding that the regularization scale matches the smallest eigenvalue of the kinematical area spectrum computed in full LQG, i.e $\Delta^2 = (4 \sqrt{3} \pi \gamma)^{3}$ where $\gamma$ is the Immirzi parameter. However, it worths emphasizing that this choice introduces an ad hoc dependency on the Immirizi parameter. Instead of introducing the Immirzi parameter, it is simpler to think of $\lambda$ as the free parameter of the polymer regularization in LQC.
}

The polymer quantization represents directly the operator $\widehat{e^{i\lambda b}}$ and consider polynomial observables of that basic operator. The regularization scale $\lambda$ is kept finite and we do not have access to the operator $\hat{b}$ anymore\footnotemark.
\footnotetext{
Moreover, the polymer representation of the algebra of observables and the associated scalar product are not continuous in the parameter $\lambda$, which means that the limit $\lambda\rightarrow 0$ in which one could hope to retrieve the operator  $\hat{b}$ is ill-defined.
}
In this context, since the scalar constraint $H$ cannot be expressed anymore in terms of the canonical variable $b$ and should be instead expressed in terms of its exponentiated version, it has to be appropriately regularized prior to quantization. This step introduces a layer of arbitrariness, since the regularization scheme is not unique. For instance, one source of ambiguity is that the observable $b$ will be replaced by an appropriate polynomial in $e^{i \lambda b}$ which does reduce to $b$ in the formal limit $b  \ll \lambda^{-1}$. Choosing the degree of this polynomial corresponds to the arbitrary choice of a spin representation when regularizing the curvature in terms of holonomy operator in the loop quantum cosmology jargon \cite{BenAchour:2016ajk}. In order to restrict these ambiguities in the construction of the theory, one can demand that the classical conformal structure, with the $\sl(2,\R)$ structure reviewed above, be preserved both by the classical regularization and in the resulting quantum theory.

\medskip

It turns out that a natural way to introduce the polymer regularization with the scale $\lambda$ is to perform a non-linear canonical transformation from the standard FLRW cosmology phase space reviewed in the previous section  \cite{BenAchour:2017qpb}. Indeed we introduce a new pair of canonical variables,
\be
\label{newvar}
B_{\lambda} (b)= \frac{\tan{\left( \lambda b\right)}}{\lambda}
\,,\qquad
\cV_{\lambda} (b,v)= v \cos^2{(\lambda b)}\,,
\ee
which satisfy $\{ B_{\lambda}, \cV_{\lambda}\} = 1$ as soon as $\{b,v\}=1$.
Notice that $v$ and $\cV_{\lambda}$ both naturally run in $\R_{+}$, but $B_{\lambda}$ runs in the whole real line $\R$ while $b$ remains bounded, $|b|\le\f\pi{2\lambda}$.

We now assume that the variables $(B_{\lambda}, \cV_{\lambda})$ follows the dynamics of standard FLRW cosmology, with the classical Hamiltonian constraint as in \eqref{Hclass}:
\be
\label{nonlinearFRW}
\cH_{\lambda}
=
\frac{\pi^2_{\phi}}{2 \cV_{\lambda}} - \frac{\ka^2}{2} \cV_{\lambda} B_{\lambda}^2
\,,
\ee
and dilatation generator $\cC_{\lambda}=B_{\lambda}\cV_{\lambda}$. We now switch back to the original variables $(b,v)$. The dilatation generator now reads:
\be
\label{regCVH1}
\cC_{\lambda} = v \frac{\sin{(2 \lambda b)}}{2\lambda}
\,,
\ee
while the classical Hamiltonian constraint becomes:
\be
\label{Hilqc}
\cH_{\lambda}[N]
=\f N2 \left(
\frac{\pi^2_{\phi}}{v \cos^2{(\lambda b)}} - \ka^2 v \frac{\sin^2{\left( \lambda b\right)}}{\lambda^2}
\right)
\,.
\ee
By construction, the CVH structure  is preserved, 
\begin{align}
\label{LQC-CVH}
\{\cV_{\lambda},\cH_{\lambda}\} &=\cC_{\lambda}
\,,\\
\{\cC_{\lambda},\cH_{\lambda}\} &=-\cH_{\lambda}
\,,\\
\{\cC_{\lambda},\cV_{\lambda}\} & =+\cV_{\lambda}
\,
\end{align}
with the same $\sl(2,\R)$ Lie algebra structure. This is our new version for an effective LQC dynamics. Since the Hamiltonian is entirely expressed in terms of $e^{i\lambda b}$, one can proceed to its polymer quantization. Before moving on to the quantization of the theory, let us make some important remarks at the classical level.

\medskip

First, one checks that in the limit where the of large volume, or equivalently small energy desnity and therefore small Hubble rate, i.e $ \lambda b \ll 1$, we come back to classical FLRW cosmology, with the three observables $\cV_{\lambda}$, $C_{\lambda}$ and $\cH_{\lambda}$ becoming back $v$, $C=vb$ and $H$ as given by \eqref{Hclass}.
Then one sees, in the regularized Hamiltonian \eqref{Hilqc},  the standard LQC squared sine correction to the gravitational sector, with the classical $vb^2$ term regularized as $v\sin^2(\lambda b)/\lambda^2$,  as well as an additional $b$-dependent correction to the inverse volume term which is absent in the standard treatment. While this new correction does not coincide with previous regularization of the inverse volume term \cite{Bojowald:2001vw}, it could originate from using a covariant version of the fluxes to write down the inverse volume term. Indeed, the covariant version of the fluxes operators carries information on the extrinsic curvature, which is usually ignored in standard LQC. Whether or not our new correction descend from this remains to be checked. We point also that similar effective corrections to the inverse volume term have been introduced in some treatment of gravitational collapse, although motivated by different considerations, such as the closure of the Dirac's hypersurface deformation algebra \cite{Campiglia:2016fzp}.

Second, it is worth emphasizing that  the sine square regularization, $v\sin^2(\lambda b)/\lambda^2$ in \eqref{Hilqc}, is not constrained by our symmetry criteria to preserve the $\sl(2,\R)$ algebra. In standard LQC, this regularization corresponds to the choice of spin $j=\f12$ in the holonomy regularization of the curvature. One is however free to choose another spin to regularize the Hamiltonian \cite{BenAchour:2016ajk}. The same ambiguity applies to the present framework and one can choose a different regularization function for the gravitational part of the Hamiltonian. Nevertheless, as shown in \cite{BenAchour:2019ywl}, preserving the $\sl(2,\mathbb{R})$ structure of the CVH algebra imposes that the regularization of the inverse volume in the matter part of the Hamiltonian is fixed in terms of the regularization chosen for the gravitational part of the Hamiltonian. The differential equations relating the two are given in  \cite{BenAchour:2019ywl}.

Third, we would like to stress that the new Hamiltonian  \eqref{Hilqc} is actually very close to the standard LQC case. We can actually map it to the usual LQC Hamiltonian by a mere re-definition of the lapse by  a $b$-dependent factor and a rescaling of the regularization scale $\lambda$ by 2:
\be
\cH_{\lambda}[N]
=\f N{2\cos^2{(\lambda b)}} \left(
\frac{\pi^2_{\phi}}{v} - \ka^2 v \frac{\sin^2{\left( 2\lambda b\right)}}{(2\lambda)^2}
\right)
\ee
Hence, we expect the dynamics of the standard and new version of LQC to  be qualitatively equivalent.
The main difference lies in the scaling properties of the volume and the scalar constraint. Indeed, by preserving the CVH algebra, the regularization scale $\lambda$ remains unaffected by scale transformation, while scale transformations in the standard treatment typically change the Immirzi parameter.
Here the regularized volume $ \cV_{\lambda}$ and regularized curvature $B_{\lambda}$ get simply rescaled under the flow generated by $\cC_{\lambda}$,
\be
e^{\eta\{ \cC_{\lambda}, \cdot\}} \triangleright \cV_{\lambda} = e^{\eta} \cV_{\lambda}
\,, \quad
e^{\eta\{ \cC_{\lambda}, \cdot\}} \triangleright B_{\lambda} = e^{-\eta} B_{\lambda}
\,,
\ee
while the transformation laws for the actual volume $v$ and physical curvature $b$ mix them together in order to leave the regularization scale $\lambda$ invariant.
Moreover, the scaling of the Hamiltonian scalar constraint is also straightforward,
\be
e^{\eta\{ \cC_{\lambda}, \cdot\}} \triangleright \cH_{\lambda} = e^{-\eta} \cH_{\lambda}
\,,
\ee
which implies that the scale transformations do not change the constraint $\cH_{\lambda}=0$ for physical state. This anticipates the fact that scales transformation will be now implemented as unitary operators at the quantum level in this new construction.
Finally, we point out that the deparametrized dynamics, using the scalar field $\phi$ as clock, consists once again simply in scale transformations generated by $\pi_{\phi}=\ka\cC_{\lambda}$.

\section{Effective Bouncing Dynamics}

\subsection{Modified Friedmann Equations}

Let us now extract  the modified cosmological dynamics of this new LQC model.
Setting the lapse $N=1$, we use the Hamilton equations of motion to describe the evolution of the volume and extrinsic curvature in terms of the cosmic time $\tau$, $\rd_{\tau}v=\{v,\cH_{\lambda}\}$ and $\rd_{\tau}b=\{b,\cH_{\lambda}\}$. The cosmological trajectory is best described by the resulting modified Friedmann equation. Let us first introduce the matter density $\rho$,
\be
\rho\equiv \f{\pi_{\phi}^2}{2a^6V_{\circ}}=\f{\pi_{\phi}^2}{2v^2},
\ee
 as well as the rescaled matter density $\tilde{\rho}$ and a critical density $\rho_c$, given by:
\begin{align}
\label{rhotilde}
\tilde{\rho}
&= \f{\pi_{\phi}^2}{2\cV^2} = \frac{\rho}{ \cos^4{\lambda b} } = \frac{4\rho}{\left(1+ \sqrt{1 - \frac{\rho}{\rho_c}}\right)^2} \,, \\
\rho_c &
=\f{\ka^2}{8\lambda^2}
= \frac{3\pi G}{2 \lambda^2}
\,.
\end{align}
We can also introduce the associated matter pressure defined as $
P=-\pp_{v}(v\rho)$
which satisfies, by definition, the equation of state $P=\rho$ for a  scalar field. Using these definitions, we get the modified Friedmann equations for the Hubble rate $\mathbb{H}\equiv \dot{v}/(3v)$ which read:
 \begin{align}
 \label{Friedmod1}
\mathbb{H}^2 
& =
\frac{8\pi G}{3} \rho \left[ 1 - \frac{\tilde{\rho}}{4\rho_c}\right]^2
\\
 \label{Friedmod2}
\dot{\mathbb{H}} & = - 4\pi G \left( \rho + P\right) \left( 1- \frac{\tilde{\rho}}{4\rho_c}\right) \left[ 1 - \frac{\tilde{\rho}}{4\rho_c} - \frac{\tilde{\rho}}{2\rho_c \sqrt{1- \frac{\rho}{\rho_c}}} \right]
\end{align}
These two equations can be combined to check that the classical continuity equation is still valid
\be
\dot{\rho} + 3H \left( \rho + P\right) =0
\ee
Going backwards in time, the density $\rho$ increases and reaches the critical density at some point, $\rho (\tau_c)= \rho_c$ and $\tilde{\rho}(\tau_{c})=4\rho_{c}$. At that time,  the Hubble factor vanishes,  $\mathbb{H}(\tau_c)=0$ while its first derivative behaves as
\be
 \dot{H}\big{|}_{\rho\rightarrow\rho_{c}} \sim 16\pi G \left( \rho + P\right) \frac{\rho}{\rho_c} 
 \ee
 which signals that the evolution of the volume freezes and the physical volume $v$ bounces. 

This means that, while the modified Friedmann equation is different from the one arising from the standard version of LQC, more precisely
\beq
\textrm{standard LQC\,:}&&
\mathbb{H}^2 = \frac{8\pi G}{3} \rho \left( 1 - \frac{{\rho}}{ \rho_c}\right)
\,,
\\
\textrm{new $\sl_{2}(\R)$ LQC\,:}&&
\mathbb{H}^2 \underset{\rho\sim\rho_{c}}\sim \frac{8\pi G}{3} \rho \left( 1 - \frac{{\rho}}{ \rho_c}\right)^2
\,,
\nn\\
&&\mathbb{H}^2 \underset{\rho\sim0}\sim \frac{8\pi G}{3} \rho \left( 1 - \frac{{\rho}}{ 2\rho_c}\right)\,,
\nn
\eeq
the resulting phenomenology of this new version of LQC fits with the standard LQC prediction of singularity resolution into a big bounce.

Here, we have discussed the classical dynamics induced by the modified Hamiltonian constraint \eqref{Hilqc}. It is nevertheless important to underline that the  those classical trajectories can be shown to be a very good approximation of the dynamics even in the deep Planckian regime. Indeed, in the effective quantum mechanics formalism developed for instance in \cite{Bojowald:2007bg,Bojowald:2014uaa}, the  $\SL(2,\R)$ symmetry protects the theory at the quantum level in the sense that the spread of the wave-function and higher order moments  only imply negligible higher order corrections and do not affect the leading order dynamics for sufficiently semi-classical states at late time (see \cite{BenAchour:2019ywl} for more details).

\subsection{Cosmological Bounce from Self-Duality}

Another much simpler method to derive the effective bounce dynamics for the modified Hamiltonian \eqref{Hilqc} is to use the essential map between this new version of LQC and standard FLRW cosmology, as given by the canonical transformation \eqref{newvar}, and the resulting relation between the 3d volume $v$ and its regularized version $\cV_{\lambda}$.

Using that the dilation generator is defined by $\lambda\cC_{\lambda}=v\cos \lambda b\sin\lambda b$, we obtain the key identity for the volume
\be
\label{vV}
v
=
v\cos^2 \lambda b+v\sin^2\lambda b
=
\cV_{\lambda}+\f{\lambda^2\cC_{\lambda}^2}{\cV_{\lambda}}
\,,
\ee
where $\cC_{\lambda}=\ka^{-1}\pi_{\phi}$ is a constant of motion given directly by the matter energy-momentum once we enforce the Hamiltonian constraint $\cH_{\lambda}=0$.

The regularized volume $\cV_{\lambda}$ evolves according to classical FLRW cosmology: it ranges from the singularity at $\cV_{\lambda}\rightarrow 0^+$ and  to  large volumes in the semi-classical regime $\cV_{\lambda}\rightarrow\infty$. We have two types of evolution:  contracting trajectories for which $\cV_{\lambda}$ evolves from $\infty$ down to the singularity, and expanding trajectories for which the volume grows 0 to $\infty$.

As we can see from \eqref{vV}, the volume $v$ mixes $\cV_{\lambda}$ and its inverse, thus mixing the two, contracting and expanding, branches. The bounce comes from the fact the key formula \eqref{vV} implies that the volume $v$ can never vanish as $\cV_{\lambda}$ evolves in $\R^+$ and its minimal allowed  value gives the  critical value $v_{c}$ for the volume at the bounce.
More precisely, the singularity  $\cV_{\lambda}\rightarrow 0^+$ is sent back to large volumes $v\rightarrow \infty$. Then, as $\cV_{\lambda}$ grows, $v$ decreases. And when $\cV_{\lambda}$ reaches the critical value $\cV_{\lambda}^c=\ka^{-1}\lambda\pi_{\phi}$, the volume $v$ reaches its minimal value $v_{c}=2\ka^{-1}\lambda\pi_{\phi}$. This is the bounce. Finally, as $\cV_{\lambda}$ grows from this critical value to infinity, the physical volume $v$ also grows back to large volumes.

\medskip

It is remarkable that the trans-Planckian regime for $\cV_{\lambda}$ from 0  to its critical value $\cV_{\lambda}^c$ is mapped to the entire contracting phase for the physical volume $v$ from $+\infty$ to its minimal value at the bounce. The singularity as $\cV_{\lambda}\rightarrow0$ is thus mapped to a semi-classical regime in LQC. 

This comes from the self-duality of the equation \eqref{vV}, which is  invariant under the exchange of small volumes and large volumes:
\be
v
=
\cV_{\lambda}+\f{(\cV_{\lambda}^c)^2}{\cV_{\lambda}}
\,,\qquad
\cV_{\lambda}\,\longleftrightarrow\, \f{(\cV_{\lambda}^c)^2}{\cV_{\lambda}}
\,.
\ee
This self-duality is reminiscent of the target space duality (T-duality) of string theory, where the compactification radius $R$ leads to dual modes with frequencies proportional to $R$ and to $R^{-1}$. In our framework and more generally in the polymer quantization scheme, the compactification occurs when going from the conjugate variable $B\in\R$ to the periodic conjugate variable $b$ with period $2\pi\lambda^{-1}$ depending on the polymer length scale.
From this perspective, the key relation \eqref{vV} and its self-duality are a direct consequences of the new regularization of the phase space preserving the $\sl(2,\R)$ structure.

Such a self-duality of loop quantum gravity models has also been noticed in effective models of quantum black holes \cite{modesto} and can be traced back more generally to the introduction of a ``zero-point length'' regularizing quantum field propagators with quantum gravity effects \cite{pad}.
Then, one can wonder if this T-duality could extend to a UV/IR mixing for the dynamics of cosmological inhomogeneties or for full loop quantum gravity, as it is a typical feature of effective non-commutative field theories introducing a universal minimal length scale (e.g. the Planck length in quantum gravity phenomenology)  through quantum group symmetries.

\section{$\sl(2,\mathbb{R})$ polymer quantization}

Turning to the quantum theory, there are two natural methods of quantization: the polymer quantization as in standard LQC or a quantization directly in terms of $\SL(2,\R)$ representations as proposed in \cite{BenAchour:2017qpb}.

On the one hand, in the polymer quantization, one splits the gravitational degrees of freedom $(b,v)$ and the matter field $(\phi,\pi_{\phi})$. It proceeds to raising $v$ and $\exp[i\lambda b]$ to quantum operators, and the volume operator $\hat{v}$ then has a discrete spectrum. The Hamiltonian constraint operator, defined as the sums of the gravitational term plus the matter contribution, then realizes the coupling between geometry and matter at the quantum level.

On the other hand, the  $\SL(2,\R)$ quantization proceeds to a direct quantization of the CVH observables. The $\sl(2,\R)$ generators $j_{z}, k_{x},k_{y}$, whose definition \eqref{sl2CVH} mixes the geometry and matter sectors, become Hermitian operators in an irreducible unitary representation of $\SL(2,\R)$, such that its Casimir is given by the matter energy-momentum reflecting the classical Casimir equation \eqref{sl2Casimir} as shown in \cite{BenAchour:2017qpb,BenAchour:2019ywl}. We do not have access to the volume $v$, which is not an elementary operator in this framework, and work instead with the generator  $\hat{j}_{z}$, which is a combination of the volume and Hamiltonian constraint and which acquires a discrete spectrum.

These two quantization scheme are subtly different and leads to different Wheeler-de-Witt equations for the quantum cosmic state. Here we focus on the polymer quantization and wish to address the fate of quantization ambiguities in light of the $\SL(2,\R)$ symmetry. For the interested reader, details on both quantization and a comparative study of the resulting quantum theory can be found in \cite{BenAchour:2019ywl}.

\medskip


The polymer quantization quantizes the geometry degrees of freedom independently from the matter field. It focuses on the observables $v$ and $e^{\pm i \lambda b}$ for the gravitational sector. The quantum commutation relations read:
\be
\label{commutrel}
\big{[}\widehat{v}, \,\widehat{U}_{\pm \lambda} \big{]}
=
\pm\lambda\,  \widehat{U}_{\pm \lambda}
\,,
\quad
\big{[}\widehat{\phi}, \,\widehat{\pi}_{\phi}\big{]}
=
i \mathds{1}
\,,
\ee
where $\widehat{U}_{\pm \lambda} \equiv \widehat{e^{\pm i \lambda b}}$ are the basic operators shifting the volume. If we introduce eigenstates $|v\ra$ for the volume operator, then $\widehat{U}_{\pm \lambda}$ increments the volume by $\pm \lambda$:
\be
\widehat{v}\,|v\ra
=
v\,|v\ra
\,,\quad
\widehat{U}_{\pm \lambda}
\,|v\ra
=
|v\pm\lambda\ra
\,.
\ee
The polymer quantization uses a scalar product different from the Shr\"odinger representation, which leads to an alternative Hilbert space. This scalar product reflects the inherent granularity of the geometry:
\be
\label{scal}
\la v | v' \ra_{{pol}} = \delta_{v,v'}
\,,
\ee
where the symbol $\delta_{v,v'}$ stands for the Kronecker-$\delta$ and not the distribution $\delta(v-v')$.
As a consequence, infinitesimal shifts of the volume are not allowed and $v$ can only change in finite steps multiple of the regularization scale $\lambda$.

This leads to a non-separable Hilbert space ${\mathbf H}^{{pol}}_g$ for gravitational states, spanned by orthonormal vectors $|v\ra$ with $v \in \mathbb{R}$. This polymer Hilbert space is a highly reducible representation of the algebra generated by the operators $\widehat{v}$ and $\widehat{U}_{\pm \lambda}$. It decomposes as the direct sum of superselection sectors ${\mathbf H}^{\eps}_g$ with $\eps\in[0,1[$ such that the spectrum of $\widehat{v}$ on ${\mathbf H}^{\eps}_g$ is $(n+\eps)\lambda$ with $n\in\Z$. We write $|n\ra_{\eps}$ short for the state $|v=(n+\eps)\lambda\ra$.
A wave function $\Psi$ in ${\mathbf H}^{\eps}_g$ then decomposes in this basis as
\begin{align}
\Psi (v) = \sum_{n \in \mathbb{Z}} \psi_{n} | n \ra_{\eps} \quad \text{such that} \,\,  \sum_{n \in \Z} |\psi_{n}|^2 <+ \infty
\,.
\end{align}
The matter sector is quantized in the standard Schr\"odinger representation. The resulting Hilbert space of quantum cosmology states is the tensor product of the matter Hilbert space and the polymer Hilbert space, ${\mathbf H}_m \otimes {\mathbf H}^{{pol}}_g$, consisting in $L^2$ wave-functions  $\Psi(\phi, v)$,
\begin{align}
\sum_{v \in \mathbb{R}} \int \rd\phi\, |\Psi(\phi, v)|^2 < + \infty
\,.
\end{align}
The basic operators act on  wave-functions $\Psi(\phi, v)$ as:
\begin{align}
\label{exp}
\widehat{U}_{\pm \lambda} \, \Psi(\phi, v) &=  \Psi(\phi, v \mp \lambda) \\
 \widehat{v} \,\Psi(\phi, v) & = v\, \Psi(\phi, v) \\
\widehat{\pi}_{\phi} \, \Psi(\phi, v) &= - i \frac{\partial}{\partial\phi}\;  \Psi(\phi, v ) \\
\widehat{\phi} \, \Psi(\phi, v) & = \phi \;  \Psi(\phi, v )
\end{align}
The Fourier transform of a state in a given sector of ${\mathbf H}_m \otimes {\mathbf H}^{{\eps}}_g$ for fixed $\eps$ reads:
\be
\label{qstate}
\Psi=\sum_{n\in\Z}\psi_{n}(\phi)\,|n\ra_{\eps}
\,,\quad
\psi_{n}(\phi)=\int_{\R}\rd k\, \widetilde{\psi}_{n}(k)e^{ik\ka\phi}
\,,
\ee
where the factor $\ka$ ensures that the exponent of the Fourier modes of the scalar field are dimensionless.
We will focus here on the sector\footnotemark{} $\eps=0$.
\footnotetext{
Representations of $\SL(2,\R)$ actually allow $\eps=0$ and $\eps=\f12$. To get all possible real values, $\eps\in[0,1[$, one uses representations of its universal cover $\widetilde{\SL(2,\R)}$.
}

\medskip

Now within this polymer representation, we would like to quantize the Hamiltonian constraint $\cH_{\lambda}$ and the dilatation generation $\cC_{\lambda}$ which generates the deparametrized evolution of the geometry in terms of the scalar field. Here we show that requiring that the $\sl(2,\R)$ symmetry realized by the CVH algebra of observables be preserved at the quantum level and represented without anomaly on the polymer Hilbert space fixes the factor-ordering ambiguities.
We separate the matter contribution from the pure geometry term in the Hamiltonian constraint \eqref{Hilqc}, explicitly $\cH_{\lambda}=\cH_{\lambda}^g+\cH_{\lambda}^m$ with:
\be
\cH_{\lambda}^g
=
 - \ka^2 v \frac{\sin^2{\left( \lambda b\right)}}{\lambda^2}
\,,\quad
\cH_{\lambda}^m
=
\frac{\pi^2_{\phi}}{v \cos^2{(\lambda b)}}
\,.
\ee
We drop the index $\lambda$ to simplify the notations.
The same way that the three observables, $\cV,\cC, \cH$ form a closed Poisson algebra, the Poisson brackets of the three observables in the purely gravitational sector $\cV,\cC, \cH^g$ also lead to a closed $\sl_{2}$ Lie algebra,
\be
\{\cV,\cH^g\}=\cC
\,,\quad
\{\cC,\cH^g\}=\cH^g
\,,\quad
\{\cC,\cV\}=\cV
\,,
\ee
but with a vanishing Casimir:
\be
\mathfrak{C}_{\sl_{2}}^g
=
-\bigg{[}
\cC^2+\f2{\ka^2}\cH^g\cV
\bigg{]}
=0
\,,
\ee
corresponding to the no-matter case $\pi_{\phi}=0$ of the full CVH algebra \eqref{sl2Casimir}.
Moving to the quantum theory, we seek operators satisfying the commutation relations:
\begin{align}
\label{sl2R-puregravity}
[\widehat{\cV}, \widehat{\cH}^g]  = i \widehat{\cC}
\,,\quad
[\widehat{\cC}, \widehat{\cH}^g]  = - i \widehat{\cH}^g
\,,\quad
[\widehat{\cC}, \widehat{\cV}]  = i \widehat{\cV}
\,.
\end{align}
It turns out that a special choice of factor ordering allows to realize this $\sl_{2}$ algebra:
\begin{align}
\widehat{\cV}\, \Psi
& = \frac{1}{4}  \sqrt{|v|} \left( \widehat{U}_{+2\lambda}  + \widehat{U}_{-2\lambda}  + 2 \right) \sqrt{|v|}
\,  \Psi
\,,
\\
\widehat{\cC}\, \Psi
& =
\frac{-i}{4\lambda } \sqrt{|v|} \left(\widehat{U}_{+2\lambda}  - \widehat{U}_{-2\lambda}  \right) \sqrt{|v|}\,  \Psi
\,, \\
\label{h}
\widehat{\cH}^g\, \Psi
& =
\frac{\ka^2}{8\lambda^2 } \sqrt{|v|} \left( \widehat{U}_{+2\lambda}  + \widehat{U}_{-2\lambda}   - 2 \right) \sqrt{|v|}
\,  \Psi
\,.
\end{align}
The Casimir of the $\sl(2,\mathbb{R})$ algebra satisfies exactly the classical relation without quantum corrections:
\begin{align}
\widehat{\mathfrak{C}}^g_{\mathfrak{sl}_{2}}
= 
-\bigg{[} 
\widehat{\cC}\,{}^2+\f1{\ka^2}\big{(}\widehat{\cH}^{g}\widehat{\cV}+\widehat{\cV}\widehat{\cH}^{g}\big{)}
\bigg{]}
=0
\end{align}
Therefore this choice of factor-ordering implies that the gravitational sector lives in a null representation of the $\sl(2,\mathbb{R})$ Lie algebra.

To investigate the dynamics of the theory and selection of the physical state, we choose the lapse function $N = \cV$ in order to sidestep the issue of the inverse volume factor in the matter term of the Hamiltonian. This gives the following Hamiltonian constraint:
\begin{align}
\label{hv}
\cH[\cV]=
\f{\pi_{\phi}^2}{2\ka^2}+\cV\cH^g
\,\,\rightarrow\,\,
\widehat{\cH[\cV]}
&= \frac{\widehat{\pi}^2_{\phi}}{2 \ka^2} + \frac{1}{2} \left(  \widehat{\cV}\widehat{\cH}^g+    \widehat{\cH}^g \widehat{\cV}  \right) 
\nn\\
&= \frac{\widehat{\pi}^2_{\phi}}{2\ka^2}  - \frac{1}{2} \widehat{\cC}\,{}^2
\,.
\end{align}
Imposing $\widehat{\cH[\cV]}=0$ to identify physical states therefore amounts to solving the equation $\widehat{\cC}=\ka^{-1}\widehat{\pi}_{\phi}$, which means that the deparametrized dynamics with respect to the scalar field time is still generated by the dilatation operator at the quantum level.

To solve the dynamics, we diagonalize the matter momentum operator by working on a fixed Fourier mode $k=\ka^{-1}{\pi}_{\phi}$ in \eqref{qstate}. Then the equation $\widehat{\cC}\,\Psi=k\,\Psi$ turns into a second order difference equation on its coefficients $\psi_{n}(k)$ in the volume eigenstate basis $|n\ra_{0}$. This difference equation implies that the zero volume state decouples from  other physical states, i.e. $\psi_0(k)=0$ for $k\neq 0$. It further implies that the Hilbert space can be decomposed into positive volume states $n>0$ and negative volume states $n<0$, which are not mixed by the dynamics. Hence, the resolution of the singularity still holds and the main results of the polymer quantization are still valid for this new version of LQC.

We refer the reader to \cite{BenAchour:2019ywl} for the explicit resolution of the Hamiltonian constraint and  the  difference equation.

\section{Conclusion}

To summarize, we have derived a new version of loop quantum cosmology (LQC) which preserves the $\sl(2,\R)$ structure and conformal symmetry of FLRW cosmology. Our results are twofold.
First, the conformal structure of the cosmological phase space provides a new criteria to restrict the quantization ambiguities when performing its canonical quantization. While some quantization ambiguities inherent to the polymer regularization such as the choice of spin in the Hamiltonian, we have shown that the factor-ordering of the volume and Hamiltonian is completely fixed by requirement of preserving the $\sl(2,\R)$ algebra at the quantum level. This should apply in principle to any quantization scheme conclusion, beyond the loop quantization investigated here.

Second, we have indeed obtained a new polymer regularization which preserves the 1d conformal symmetry of FLRW cosmology, at both classical and quantum level. In particular, the generator of the 3d scale transformations, which turns out to also generate the deparametrized dynamics of the geometry with respect to the scalar field, is realized as a Hermitian operator so that 3d scale transformations become unitary operators. This is a new feature for LQC: keeping unitary scale transformations while introducing a minimal length scale as a universal cut-off for the quantum theory. 


At the end of the day, this approach allows to recast the minimal model of LQC as an $\sl(2,\mathbb{R})$ quantum cosmology with discrete volume spectrum.
It invites further investigation of the LQC framework. For instance,  the canonical transformation \eqref{newvar} provides a shortcut to derive a $\sl(2,\R)$-consistent polymer regularization from the standard FLRW phase space. It would be interesting to re-derive it from a standard holonomy-flux regularization following the standard LQG regularization technics. The  challenge would be in particular to identify the origin of the cosine regularization of the inverse volume term.

Beyond such technical improvements, the $\SL(2,\R)$ structure was already recently used to build an elegant coarse-graining procedure based on  group coherent states \cite{Bodendorfer:2018csn}.
And our results  further opens  new lines of research such as bootstrapping quantum cosmology using the $\SL(2,\R)$ symmetry and a mapping to  conformal quantum mechanics \cite{deAlfaro:1976vlx} used in the investigation of a possible AdS${}_{2}$/CFT${}_{1}$ correspondence. It could also turn out useful to build a consistent holographic model of quantum cosmology \cite{Okazaki:2015lpa}.
Additionally, the $\sl(2,\R)$ algebra and its potential extension to a Virasoro symmetry suggests a relation with the non-linear Schr\"odinger equations describing low-dimensional condensates \cite{Lidsey:2018byv}, which could pave the way to a reformulation of LQC in terms of quantum condensates.

Let us finally stress that the $\sl(2,\R)$ structure and the corresponding polymer regularization presented here are straightforwardly generalized to the anisotropic Bianchi I cosmology, suggesting that this conformal structure can hold in more general gravitational systems and is not a consequence of the high degree of symmetry of the homogeneous and isotropic sector studied in this work.

\textit{Acknowledgements.} This work was supported by the National Science Foundation of China, Grant No.11875006 (J. BA), and by the China Postdoctoral Science Foundation  with Grant No. 212400209 (J. BA).




\begin{thebibliography}{10}

\bibitem{BenAchour:2019ywl}
J.~Ben~Achour and E.~R. Livine, ``{Protected $SL(2,\mathbb{R})$ Symmetry in
  Quantum Cosmology},''
\href{http://arXiv.org/abs/1904.06149}{{\texttt{arXiv:1904.06149}}}.

\bibitem{Wiltshire:1995vk}
D.~L. Wiltshire, ``{An Introduction to quantum cosmology},'' in {\em
  {Proceedings, 8th Physics Summer School, Canberra, Australia, Jan 16-Feb 3,
  1995}}, pp.~473--531.
\newblock 1995.
\newblock
\href{http://arXiv.org/abs/gr-qc/0101003}{{\texttt{arXiv:gr-qc/0101003}}}.
\newblock

\bibitem{Bojowald:2010cj}
M.~Bojowald, C.~Kiefer, and P.~Vargas~Moniz, ``{Quantum cosmology for the 21st
  century: A Debate},'' in {\em {Proceedings, 12th Marcel Grossmann Meeting on
  General Relativity, Paris, France, July 12-18, 2009. Vol. 1-3}},
  pp.~589--608.
\newblock 2010.
\newblock
\href{http://arXiv.org/abs/1005.2471}{{\texttt{arXiv:1005.2471}}}.
\newblock

\bibitem{Bojowald:2015iga}
M.~Bojowald, ``{Quantum cosmology: a review},'' Rept. Prog. Phys. {\bf 78}
  (2015) 023901,
\href{http://arXiv.org/abs/1501.04899}{{\texttt{arXiv:1501.04899}}}.

\bibitem{Steigl:2005fk}
R.~Steigl and F.~Hinterleitner, ``{Factor ordering in standard quantum
  cosmology},'' Class. Quant. Grav. {\bf 23} (2006) 3879--3894,
\href{http://arXiv.org/abs/gr-qc/0511149}{{\texttt{arXiv:gr-qc/0511149}}}.

\bibitem{Bojowald:2014ija}
M.~Bojowald and D.~Simpson, ``{Factor ordering and large-volume dynamics in
  quantum cosmology},'' Class. Quant. Grav. {\bf 31} (2014) 185016,
\href{http://arXiv.org/abs/1403.6746}{{\texttt{arXiv:1403.6746}}}.

\bibitem{Livine:2012mh}
E.~R. Livine and M.~Martin-Benito, ``{Group theoretical Quantization of
  Isotropic Loop Cosmology},'' Phys. Rev. {\bf D85} (2012) 124052,
\href{http://arXiv.org/abs/1204.0539}{{\texttt{arXiv:1204.0539}}}.

\bibitem{Corichi:2006qf}
A.~Corichi, T.~Vukasinac, and J.~A. Zapata, ``{Hamiltonian and physical Hilbert
  space in polymer quantum mechanics},'' Class. Quant. Grav. {\bf 24} (2007)
  1495--1512,
\href{http://arXiv.org/abs/gr-qc/0610072}{{\texttt{arXiv:gr-qc/0610072}}}.

\bibitem{Fredenhagen:2006wp}
K.~Fredenhagen and F.~Reszewski, ``{Polymer state approximations of Schrodinger
  wave functions},'' Class. Quant. Grav. {\bf 23} (2006) 6577--6584,
\href{http://arXiv.org/abs/gr-qc/0606090}{{\texttt{arXiv:gr-qc/0606090}}}.

\bibitem{Corichi:2007tf}
A.~Corichi, T.~Vukasinac, and J.~A. Zapata, ``{Polymer Quantum Mechanics and
  its Continuum Limit},'' Phys. Rev. {\bf D76} (2007) 044016,
\href{http://arXiv.org/abs/0704.0007}{{\texttt{arXiv:0704.0007}}}.

\bibitem{BenAchour:2016ajk}
J.~Ben~Achour, S.~Brahma, and M.~Geiller, ``{New Hamiltonians for loop quantum
  cosmology with arbitrary spin representations},'' Phys. Rev. {\bf D95}
  (2017), no.~8, 086015,
\href{http://arXiv.org/abs/1612.07615}{{\texttt{arXiv:1612.07615}}}.

\bibitem{Singh:2013ava}
P.~Singh and E.~Wilson-Ewing, ``{Quantization ambiguities and bounds on
  geometric scalars in anisotropic loop quantum cosmology},'' Class. Quant.
  Grav. {\bf 31} (2014) 035010,
\href{http://arXiv.org/abs/1310.6728}{{\texttt{arXiv:1310.6728}}}.

\bibitem{Corichi:2011pg}
A.~Corichi and A.~Karami, ``{Loop quantum cosmology of k=1 FRW: A tale of two
  bounces},'' Phys. Rev. {\bf D84} (2011) 044003,
\href{http://arXiv.org/abs/1105.3724}{{\texttt{arXiv:1105.3724}}}.

\bibitem{Dupuy:2016upu}
J.~L. Dupuy and P.~Singh, ``{Implications of quantum ambiguities in $k$=1 loop
  quantum cosmology: distinct quantum turnarounds and the super-Planckian
  regime},'' Phys. Rev. {\bf D95} (2017), no.~2, 023510,
\href{http://arXiv.org/abs/1608.07772}{{\texttt{arXiv:1608.07772}}}.

\bibitem{deAlfaro:1976vlx}
V.~de~Alfaro, S.~Fubini, and G.~Furlan, ``{Conformal Invariance in Quantum
  Mechanics},'' Nuovo Cim. {\bf A34} (1976)
569.

\bibitem{BenAchour:2017qpb}
J.~Ben~Achour and E.~R. Livine, ``{Thiemann complexifier in classical and
  quantum FLRW cosmology},'' Phys. Rev. {\bf D96} (2017), no.~6, 066025,
\href{http://arXiv.org/abs/1705.03772}{{\texttt{arXiv:1705.03772}}}.

\bibitem{Rovelli:1997na}
C.~Rovelli and T.~Thiemann, ``{The Immirzi parameter in quantum general
  relativity},'' Phys. Rev. {\bf D57} (1998) 1009--1014,
\href{http://arXiv.org/abs/gr-qc/9705059}{{\texttt{arXiv:gr-qc/9705059}}}.

\bibitem{Perez:2005pm}
A.~Perez and C.~Rovelli, ``{Physical effects of the Immirzi parameter},'' Phys.
  Rev. {\bf D73} (2006) 044013,
\href{http://arXiv.org/abs/gr-qc/0505081}{{\texttt{arXiv:gr-qc/0505081}}}.

\bibitem{Pioline:2002qz}
B.~Pioline and A.~Waldron, ``{Quantum cosmology and conformal invariance},''
  Phys. Rev. Lett. {\bf 90} (2003) 031302,
\href{http://arXiv.org/abs/hep-th/0209044}{{\texttt{arXiv:hep-th/0209044}}}.

\bibitem{Bojowald:2001vw}
M.~Bojowald, ``{The Inverse scale factor in isotropic quantum geometry},''
  Phys. Rev. {\bf D64} (2001) 084018,
\href{http://arXiv.org/abs/gr-qc/0105067}{{\texttt{arXiv:gr-qc/0105067}}}.

\bibitem{Campiglia:2016fzp}
M.~Campiglia, R.~Gambini, J.~Olmedo, and J.~Pullin, ``{Quantum self-gravitating
  collapsing matter in a quantum geometry},'' Class. Quant. Grav. {\bf 33}
  (2016), no.~18, 18LT01,
\href{http://arXiv.org/abs/1601.05688}{{\texttt{arXiv:1601.05688}}}.

\bibitem{Bojowald:2007bg}
M.~Bojowald, ``{Dynamical coherent states and physical solutions of quantum
  cosmological bounces},'' Phys. Rev. {\bf D75} (2007) 123512,
\href{http://arXiv.org/abs/gr-qc/0703144}{{\texttt{arXiv:gr-qc/0703144}}}.

\bibitem{Bojowald:2014uaa}
M.~Bojowald and A.~Tsobanjan, ``{Effective Casimir Conditions and Group
  Coherent States},'' Class. Quant. Grav. {\bf 31} (2014) 115006,
\href{http://arXiv.org/abs/1401.5352}{{\texttt{arXiv:1401.5352}}}.

\bibitem{modesto} 
  L.~Modesto and I.~Premont-Schwarz,
  ``Self-dual Black Holes in LQG: Theory and Phenomenology,''
  Phys.\ Rev.\ D {\bf 80}, 064041 (2009),
  \href{http://arxiv.org/abs/0905.3170}{{\texttt{arXiv:0905.3170}}}.

\bibitem{pad} 
  T.~Padmanabhan,
  ``Duality and zero point length of space-time,''
  Phys.\ Rev.\ Lett.\  {\bf 78}, 1854 (1997),
  \href{http://arxiv.org/abs/hep-th/9608182}{{\texttt{arXiv:hep-th/9608182}}}.

\bibitem{Bodendorfer:2018csn}
N.~Bodendorfer and F.~Haneder, ``{Coarse graining as a representation
  change},'' Phys. Lett. {\bf B792} (2019) 69--73,
\href{http://arXiv.org/abs/1811.02792}{{\texttt{arXiv:1811.02792}}}.

\bibitem{Okazaki:2015lpa}
T.~Okazaki, ``{Whittaker vector, Wheeler-DeWitt equation, and the gravity dual
  of conformal quantum mechanics},'' Phys. Rev. {\bf D92} (2015), no.~12,
  126010,
\href{http://arXiv.org/abs/1510.04759}{{\texttt{arXiv:1510.04759}}}.

\bibitem{Lidsey:2018byv}
J.~E. Lidsey, ``{Inflationary Cosmology, Diffeomorphism Group of the Line and
  Virasoro Coadjoint Orbits},''
\href{http://arXiv.org/abs/1802.09186}{{\texttt{arXiv:1802.09186}}}.

\end{thebibliography}

\end{document}